\newcounter{myctr}
\def\myitem{\refstepcounter{myctr}\bibfont\noindent\ifnum\themyctr>9\else\phantom{0}\fi\hangindent17pt\themyctr.\enskip}

\documentclass{ws-ijqi}

\begin{document}

\markboth{Haoyang Wu} {Quantum mechanism helps agents combat ``bad''
social choice rules}

\catchline{}{}{}{}{}

\title{Quantum mechanism helps agents combat ``bad'' \\social choice rules  }

\author{Haoyang Wu}

\address{Wan-Dou-Miao Research Lab, Suite 1002, 790 WuYi Road,\\
Shanghai, 200051, China.\\
hywch@mail.xjtu.edu.cn}

\maketitle


\begin{abstract}
Quantum strategies have been successfully applied to game theory for
years. However, as a reverse problem of game theory, the theory of
mechanism design is ignored by physicists. In this paper, the theory
of mechanism design is generalized to a quantum domain. The main
result is that by virtue of a quantum mechanism, agents who satisfy
a certain condition can combat ``bad'' social choice rules instead
of being restricted by the traditional mechanism design theory.
\end{abstract}


\section{Introduction}

Game theory is a very useful tool for investigating rational
decision making in conflict situations. It was first founded by von
Neumann and Morgenstern $^{1}$. Since its beginning, game theory has
been widely applied to many disciplines, such as economics,
politics, biology and so on. Compared with game theory, the theory
of mechanism design simply concerns the \emph{reverse} question:
given some desirable outcomes, can we design a game that produces
them?

As Serrano  $^{2}$ has described, we suppose that the goals of a
group of self-interested agents (or a society) can be summarized in
a social choice rule (SCR). An SCR is a mapping that prescribes the
social outcome (or outcomes) on the basis of agents' preferences
over the set of all social outcomes  $^{3}$. The theory of mechanism
design answers the important question of whether and how it is
possible to implement different SCRs. According to Maskin and
Sj\"{o}str\"{o}m  $^{4}$, whether or not an SCR is implementable
depends on which game theoretic solution concept is used (e.g.,
dominant strategies and Nash equilibrium). Reference 3 is a
fundamental work in the field of mechanism design. It provides an
almost complete characterization of social choice rules that are
Nash implementable.

In 1999, some pioneering breakthroughs were made in the field of
quantum games  $^{5,6}$. The game proposed by Eisert \emph{et al}
 $^{5}$ showed fascinating ``quantum advantages'' as a
result of a novel quantum Nash equilibrium. Benjamin and Hayden
 $^{7}$, Du \emph{et al}  $^{8}$, Flitney and
Hollenberg  $^{9}$ investigated multiplayer quantum Prisoner's
Dilemma. Guo \emph{et al} $^{10}$ gave a detailed review on quantum
games. As a comparison, so far the theory of mechanism design is
still investigated only by economists. To the best of our knowledge,
up to now, there is no research in the cross field between quantum
mechanics and mechanism design. Motivated by quantum games, in this
paper, we will investigate what will happen if agents can use
quantum strategies in the theory of mechanism design.

Section 2 of this paper recalls some preliminaries of mechanism
design published in Ref. 2, while Sec. 3 reformulates the Maskin's
mechanism as a physical mechanism and proves that they are
equivalent to each other. Section 4 generalizes the physical
mechanism to a quantum domain and proves that under a certain
condition, an original Nash implementable social choice rule will no
longer be implemented. Section 5 draws the conclusions.

\section{Preliminaries}
Let $N=\{1,\cdots,n\}$ be a finite set of \emph{agents} with $n\geq
2$ and $A=\{a_{1},\cdots,a_{k}\}$ be a finite set of social
\emph{outcomes}. Let $T_{i}$ be the finite set of agent $i$'s types,
and the \emph{private information} possessed by agent $i$ is denoted
as $t_{i}\in T_{i}$. We refer to a profile of types
$t=(t_{1},\cdots,t_{n})$ as a \emph{state}. Let $\mathcal
{T}=\prod_{i\in N}T_{i}$ be the set of states. At state
$t\in\mathcal {T}$, each agent $i\in N$ is assumed to have a
complete and transitive \emph{preference relation} $\succeq_{i}^{t}$
over the set $A$. We denote by
$\succeq^{t}=(\succeq_{1}^{t},\cdots,\succeq_{n}^{t})$ the profile
of preferences in state $t$. The utility of agent $i$ for outcome
$a$ in state $t$ is $u_{i}(a,t):A\times \mathcal {T}\rightarrow R$,
i.e., $u_{i}(a,t)\geq u_{i}(b,t)$ if and only if $a\succeq_{i}^{t}
b$. We denote by $\succ_{i}^{t}$ the strict preference part of
$\succeq_{i}^{t}$. Fixing a state $t$, we refer to the collection
$E=<N,A,(\succeq_{i}^{t})_{i\in N}>$ as an \emph{environment}. Let
$\varepsilon$ be the class of possible environments. A \emph{social
choice rule} (SCR) $F$ is a mapping $F:\varepsilon\rightarrow
2^{A}\backslash\{\emptyset\}$. A \emph{mechanism}
$\Gamma=((M_{i})_{i\in N},g)$ describes a message or strategy set
$M_{i}$ for agent $i$, and an outcome function $g:\prod_{i\in
N}M_{i}\rightarrow A$.

An SCR $F$ satisfies \emph{no-veto} if, whenever $a\succeq_{i}^{t}b$
for all $b\in A$ and for all agents $i$ but perhaps one $j$, then
$a\in F(E)$. An SCR $F$ is \emph{monotonic} if for every pair of
environments $E$ and $E'$, and for every $a\in F(E)$, whenever
$a\succeq_{i}^{t}b$ implies that $a\succeq_{i}^{t'}b$, there holds
$a\in F(E')$. We assume that there is \emph{complete information}
among the agents, i.e., the true state $t$ is common knowledge among
them. Given a mechanism $\Gamma=((M_{i})_{i\in N},g)$ played in
state $t$, a \emph{Nash equilibrium} of $\Gamma$ in state $t$ is a
strategy profile $m^{*}$ such that: $\forall i\in N,
g(m^{*}(t))\succeq_{i}^{t}g(m_{i},m_{-i}^{*}(t)), \forall m_{i}\in
M_{i}$. Let $\mathcal {N}(\Gamma,t)$ denote the set of Nash
equilibria of the game induced by $\Gamma$ in state $t$, and
$g(\mathcal {N}(\Gamma,t))$ denote the corresponding set of Nash
equilibrium outcomes. An SCR $F$ is \emph{Nash implementable} if
there exists a mechanism $\Gamma=((M_{i})_{i\in N},g)$ such that for
every $t\in \mathcal {T}$, $g(\mathcal {N}(\Gamma,t))=F(t)$.

Maskin  $^{3}$ provided an almost complete characterization of
social choice rules that were Nash implementable. The main results
of Ref. 3 are two theorems: (i) (Necessity) If an SCR $F$ is Nash
implementable, then it is monotonic. (ii) (Sufficiency) Let
$n\geq3$, if an SCR $F$ is monotonic and satisfies no-veto, then it
is Nash implementable. In order to facilitate the following
investigation on quantum mechanism, we briefly recall the Maskin's
mechanism as follows
 $^{2}$:

Let $\mathbb{Z}_{+}$ be the set of non-negative integers.
Considering the following mechanism $\Gamma=((M_{i})_{i\in N},g)$,
where agent $i$'s message set is $M_{i}=A\times \mathcal {T} \times
\mathbb{Z}_{+}$, we denote a typical message sent by agent $i$ by
$m_{i}=(a_{i},t_{i},z_{i})$. The outcome function $g$ is defined in
the following three rules: (1) If for every agent $i\in N$,
$m_{i}=(a,t,0)$ and $a\in F(t)$, then $g(m)=a$. (2) If $(n-1)$
agents $i\neq j$ send $m_{i}=(a,t,0)$ and $a\in F(t)$, but agent $j$
sends $m_{j}=(a_{j},t_{j},z_{j})\neq(a,t,0)$, then $g(m)=a$ if
$a_{j}\succ_{j}^{t}a$, and $g(m)=a_{j}$ otherwise. (3) In all other
cases, $g(m)=a'$, where $a'$ is the outcome chosen by the agent with
the lowest index among those who announce the highest integer.

\section{Physical mechanism}
It can be seen that in the Maskin's mechanism, a message is an
abstract mathematical notion. People usually neglect how it is
realized physically. However, the world is a physical world. Any
information must be related to a physical entity. Here we assume:

(i) Each agent has a coin and a card. The state of a coin can be
head up or tail up (denoted as $H$ and $T$ respectively).

(ii) Each agent $i$ independently chooses a strategic action
$\omega_{i}$ whether to flip his/her coin. The set of agent $i$'s
action is $\Omega_{i}=$\{\emph{Not flip}, \emph{Flip}\}. An action
$\omega_{i}\in\Omega_{i}$ chosen by agent $i$ is defined as
$\omega_{i}: \{H,T\}\rightarrow\{H,T\}$. If $\omega_{i}$=\emph{Not
flip}, then $\omega_{i}(H)=H$, $\omega_{i}(T)=T$; If
$\omega_{i}$=\emph{Flip}, then $\omega_{i}(H)=T$, $\omega_{i}(T)=H$.

(iii) The two sides of a card are denoted as Side 0 and Side 1. The
message written on the Side 0 (or Side 1) of card $i$ is denoted as
$card(i,0)$ (or $card(i,1)$).

(iv) There is a device that can measure the state of $n$ coins and
send messages to the designer.

Based on aforementioned assumptions, we reformulate the Maskin's
mechanism $\Gamma=((M_{i})_{i\in N},g)$ as a \emph{physical
mechanism} $\Gamma^{P}=((S_{i})_{i\in N},G)$, where
$S_{i}=\Omega_{i}\times C_{i}$, $C_{i}$ is agent $i$'s card set,
$C_{i}=A\times \mathcal {T} \times \mathbb{Z}_{+}\times A\times
\mathcal {T} \times \mathbb{Z}_{+}$. A typical card written by agent
$i$ is described as $c_{i}=(card(i,0),card(i,1))$, where
$card(i,0)=(a_{i},t_{i},z_{i})$, $card(i,1)=(a'_{i},t'_{i},z'_{i})$.
A physical mechanism $\Gamma^{P}=((S_{i})_{i\in N},G)$ describes a
strategy set $S_{i}$ for agent $i$ and an outcome function
$G:\prod_{i\in N}S_{i}\rightarrow A$. We shall use $S_{-i}$ to
express $\prod_{j\neq i}S_{j}$, and thus, a strategy profile is
$s=(s_{i},s_{-i})$, where $s_{i}=(\omega_{i},c_{i})\in S_{i}$ and
$s_{-i}=(\omega_{-i},c_{-i})\in S_{-i}$. A \emph{Nash equilibrium}
of $\Gamma^{P}$ played in state $t$ is a strategy profile
$s^{*}=(s^{*}_{1},\cdots,s^{*}_{n})$ such that for any agent $i\in
N$, $s_{i}\in S_{i}$, $G(s^{*}_{1},\cdots,s^{*}_{n})\succeq^{t}_{i}
G(s_{i},s^{*}_{-i})$. Figure 1 depicts the setup of a physical
mechanism. From the viewpoint of the designer, the physical
mechanism works in the same manner as the Maskin's mechanism. The
working steps of the physical mechanism are shown as follows:\\
Step 1: Nature selects a state $t\in \mathcal {T}$ and assigns $t$
to the agents. Each coin is set head up.\\
Step 2: In state $t$, if all agents agree that the social choice
rule $F$ is Pareto-inefficient (or ``\emph{bad}''), i.e., there
exist $\hat{t}\in \mathcal {T}$, $\hat{t}\neq t$, $\hat{a}\in
F(\hat{t})$ such that $\hat{a}\succeq^{t}_{i}a\in F(t)$ for every
$i\in N$, and $\hat{a}\succ^{t}_{j}a\in F(t)$ for at least one $j\in
N$, then go to Step 4.\\
Step 3: Each agent $i$ sets
$c_{i}=((a_{i},t_{i},z_{i}),(a_{i},t_{i},z_{i}))$ (where $a_{i}\in
A$, $t_{i}\in\mathcal {T}$, $z_{i}\in\mathbb{Z}_{+}$),
$\omega_{i}=$\emph{Not flip}. Go to Step 5.\\
Step 4: Each agent $i$ sets
$c_{i}=((\hat{a},\hat{t},0),(a_{i},t_{i},z_{i}))$, then chooses a
strategic action $\omega_{i}\in\Omega_{i}$ whether to flip coin
$i$.\\
Step 5: The device measures the state of $n$ coins and sends
$card(i,0)$ (or $card(i,1)$) as $m_{i}$ to the designer if coin $i$
is head up (or tail up). The designer receives the overall message
$m=(m_{1},\cdots,m_{n})$ and let the final outcome be $G(s)=g(m)$
using rule (1), (2) and (3) defined in the Maskin's mechanism. END.

\begin{figure}
\centering
\includegraphics[height=2.5in,clip,keepaspectratio]{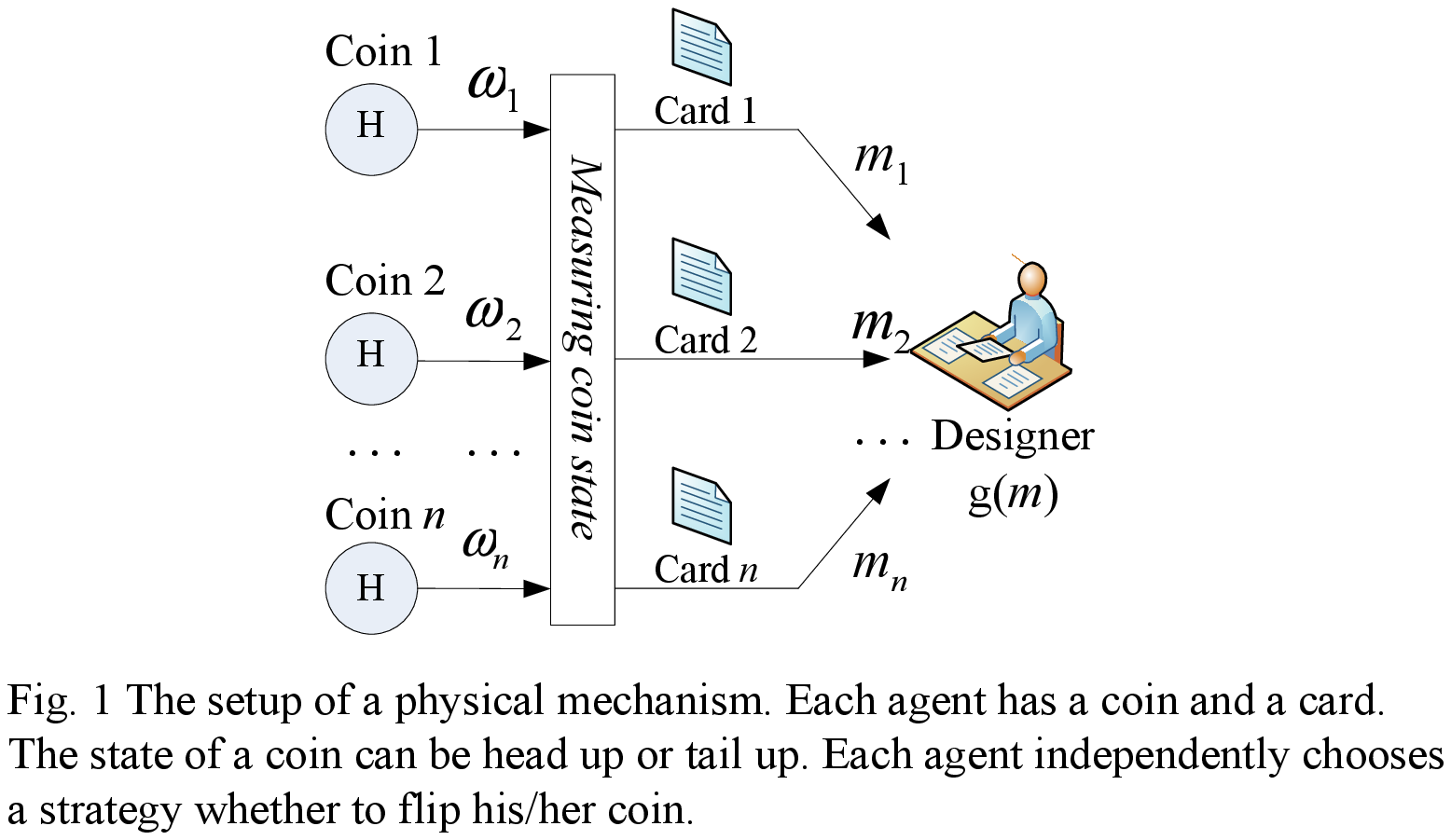}
\end{figure}

\textbf{Proposition 1:} Given an SCR $F$ and a state $t\in \mathcal
{T}$, $\mathcal{N}(\Gamma^{P},t)$ is equivalent to
$\mathcal{N}(\Gamma,t)$.

\textbf{Proof:} First, define a function
$R:\{H,T\}\rightarrow\{0,1\}$, $R(H)=0$, $R(T)=1$. For any
$s^{*}=(s^{*}_{1},\cdots,s^{*}_{n})\in\mathcal{N}(\Gamma^{P},t)$ and
$a=G(s^{*})$, if $a$ is generated by Step 4 and 5, then for each
agent $i$, let $m^{*}_{i}=card(i,R(\omega^{*}_{i}(H)))$; if $a$ is
generated by Step 3 and 5, then for each agent $i$, let
$m^{*}_{i}=card(i,0)$. Obviously,
$m^{*}=(m^{*}_{1},\cdots,m^{*}_{n})\in\mathcal{N}(\Gamma,t)$.

Next, for any
$m^{*}=(m^{*}_{1},\cdots,m^{*}_{n})\in\mathcal{N}(\Gamma,t)$, for
each agent $i$, let $s_{i}^{*}=(\omega_{i}^{*}, c_{i}^{*})$, where
$\omega^{*}_{i}$=\emph{Not flip}, $c_{i}^{*}=(m^{*}_{i},m^{*}_{i})$,
then
$s^{*}=(s^{*}_{1},\cdots,s^{*}_{n})\in\mathcal{N}(\Gamma^{P},t)$.
\quad $\square$

\begin{table}
\tbl{An example of a ``bad'' SCR  that is monotonic and satisfies
no-veto.} {\begin{tabular}{cccccc}
 \multicolumn{3}{c}{State $t_{1}$}&\multicolumn{3}{c}{State $t_{2}$}\\
 $Apple$&$Lily$ &$Cindy$ &$Apple$&$Lily$ &$Cindy$\\ \hline
 $a_{3}$&$a_{2}$&$a_{1}$ &$a_{4}$&$a_{3}$&$a_{1}$ \\
 $a_{1}$&$a_{1}$&$a_{3}$ &$a_{1}$&$a_{1}$&$a_{2}$ \\
 $a_{2}$&$a_{4}$&$a_{2}$ &$a_{2}$&$a_{2}$&$a_{3}$ \\
 $a_{4}$&$a_{3}$&$a_{4}$ &$a_{3}$&$a_{4}$&$a_{4}$ \\\hline
 \multicolumn{3}{c}{$F(t_{1})=\{a_{1}\}$}&\multicolumn{3}{c}{$F(t_{2})=\{a_{2}\}$}\\\hline
\end{tabular}}
\end{table}

\textbf{Example 1:} Let $N=\{Apple, Lily, Cindy\}$, $\mathcal
{T}=\{t_{1},t_{2}\}$, $A=\{a_{1},a_{2},a_{3},a_{4}\}$. In each state
$t\in\mathcal {T}$, the preference relations
$(\succeq^{t}_{i})_{i\in N}$ over the outcome set $A$ and the
corresponding SCR $F$ are given in Table 1. Obviously, $F$ is
monotonic and satisfies no-veto. By Maskin's theorem, $F$ is Nash
implementable. The SCR $F$ is ``bad'' from the viewpoint of the
agents because in state $t=t_{2}$, all agents unanimously prefer a
Pareto-efficient outcome $a_{1}\in F(t_{1})$: for each agent $i$,
$a_{1}\succ^{t_{2}}_{i}a_{2}\in F(t_{2})$. Therefore when the true
state is $t_{2}$, the physical mechanism enters Step 4.

Since every agent prefers $a_{1}$ to $a_{2}$ in state $t_{2}$, it
seems that for each agent $i$, $(\hat{a},\hat{t},0)=(a_{1},t_{1},0)$
should be a unanimous $card(i,0)$, and ``\emph{Not flip}'' be the
same strategic action. As a result, the outcome $a_{1}$ may be
generated by rule (1). However, $Apple$ has an incentive to
unilaterally deviate from $(a_{1},t_{1},0)$ to $(a_{4},*,*)$ by
flipping her coin, since $a_{1}\succ^{t_{1}}_{Apple}a_{4}$,
$a_{4}\succ^{t_{2}}_{Apple}a_{1}$; $Lily$ also has an incentive to
unilaterally deviate from $(a_{1},t_{1},0)$ to $(a_{3},*,*)$ by
flipping her coin, since $a_{1}\succ^{t_{1}}_{Lily}a_{3}$,
$a_{3}\succ^{t_{2}}_{Lily}a_{1}$. $Cindy$ has no incentive to
deviate from $(a_{1},t_{1},0)$ because $a_{1}$ is her top-ranked
outcome in two states. Therefore,
$c_{Apple}=((a_{1},t_{1},0),(a_{4},*,*))$,
$c_{Lily}=((a_{1},t_{1},0),(a_{3},*,*))$,
$c_{Cindy}=((a_{1},t_{1},0),(a_{1},t_{1},0))$.

Note that either $Apple$ or $Lily$ can certainly obtain her expected
outcome only if just one of them flips her coin and deviates from
$(a_{1},t_{1},0)$ (If this case happens, rule (2) will be
triggered). But this condition is unreasonable, because all agents
are rational, nobody is willing to give up and let the others
benefit. Therefore, both $Apple$ and $Lily$ will flip their coins
and deviate from $(a_{1},t_{1},0)$. As a result, rule (3) will be
triggered. Since $Apple$ and $Lily$ both have a chance to win the
integer game, the winner is uncertain. Consequently, the final
outcome is uncertain between $a_{3}$ and $a_{4}$, denoted as
$a_{3}/a_{4}$.

To sum up, although every agent prefers $a_{1}$ to $a_{2}$ in state
$t=t_{2}$, $a_{1}$ cannot be generated in Nash equilibrium. Indeed,
the Maskin's mechanism makes the Pareto-inefficient outcome $a_{2}$
be Nash implementable in state $t=t_{2}$. The underlying reason is
just the same as what we have seen in the well-known Prisoner's
Dilemma, i.e., the individual rationality is in conflict with the
group rationality. In this sense, the agents cannot combat the
``bad'' SCR under the classical circumstance.

\section{Quantum mechanism}
In 2007, Flitney and Hollenberg $^{9}$ investigated Nash equilibria
in $n$-player quantum Prisoner's Dilemma. Following their
procedures, we define:
\begin{equation*}
\hat{\omega}(\theta,\phi)\equiv \begin{bmatrix}
  e^{i\phi}\cos(\theta/2) & i\sin(\theta/2)\\
  i\sin(\theta/2) & e^{-i\phi}\cos(\theta/2)
\end{bmatrix},
\end{equation*}
$\hat{\Omega}\equiv\{\hat{\omega}(\theta,\phi):\theta\in[0,\pi],\phi\in[0,\pi/2]\}$,
$\hat{J}\equiv\cos(\gamma/2)\hat{I}^{\otimes
n}+i\sin(\gamma/2)\hat{\sigma_{x}}^{\otimes
  n}$,
where $\gamma$ is an entanglement measure, and
$\hat{I}\equiv\hat{\omega}(0,0)$,
$\hat{D}_{n}\equiv\hat{\omega}(\pi,\pi/n)$,
$\hat{C}_{n}\equiv\hat{\omega}(0,\pi/n)$.

In order to generalize the physical mechanism to a quantum domain,
we revise the assumptions (i) and (ii) of the physical mechanism as
follows:

1) Each agent $i$ has a quantum coin $i$ (qubit) and a classical
card $i$. The basis vectors $|C\rangle\equiv(1,0)^{T}$,
$|D\rangle\equiv(0,1)^{T}$ of a quantum coin denote head up and tail
up respectively.

2) Each agent $i$ independently performs a local unitary operation
on his/her own quantum coin. The set of agent $i$'s operation is
$\hat{\Omega}_{i}=\hat{\Omega}$. A strategic operation chosen by
agent $i$ is denoted as $\hat{\omega}_{i}\in\hat{\Omega}_{i}$. If
$\hat{\omega}_{i}=\hat{I}$, then
$\hat{\omega}_{i}(|C\rangle)=|C\rangle$,
$\hat{\omega}_{i}(|D\rangle)=|D\rangle$; If
$\hat{\omega}_{i}=\hat{D}_{n}$, then
$\hat{\omega}_{i}(|C\rangle)=|D\rangle$,
$\hat{\omega}_{i}(|D\rangle)=|C\rangle$. $\hat{I}$ denotes
``\emph{Not flip}'', $\hat{D}_{n}$ denotes ``\emph{Flip}''.

Based on aforementioned amendments, we generalize the physical
mechanism $\Gamma^{P}=((S_{i})_{i\in N},G)$ to a \emph{quantum
mechanism} $\Gamma^{Q}=((\hat{S}_{i})_{i\in N},\hat{G})$, which
describes a strategy set $\hat{S}_{i}=\hat{\Omega}_{i}\times C_{i}$
for each agent $i$ and an outcome function $\hat{G}:\otimes_{i\in
N}\hat{\Omega}_{i}\times\prod_{i\in N}C_{i}\rightarrow A$. We shall
use $\hat{S}_{-i}$ to express $\otimes_{j\neq
i}\hat{\Omega}_{j}\times\prod_{j\neq i}C_{j}$, and thus, a strategy
profile is $\hat{s}=(\hat{s}_{i},\hat{s}_{-i})$, where
$\hat{s}_{i}\in\hat{S}_{i}$ and $\hat{s}_{-i}\in\hat{S}_{-i}$. A
\emph{Nash equilibrium} of a quantum mechanism $\Gamma^{Q}$ played
in state $t$ is a strategy profile
$\hat{s}^{*}=(\hat{s}^{*}_{1},\cdots,\hat{s}^{*}_{n})$ such that for
any agent $i\in N$, $\hat{s}_{i}\in\hat{S}_{i}$,
$\hat{G}(\hat{s}^{*}_{1},\cdots,\hat{s}^{*}_{n})\succeq^{t}_{i}
\hat{G}(\hat{s}_{i},\hat{s}^{*}_{-i})$. Figure 2 depicts the set-up
of a quantum mechanism. Its working steps are shown as follows:

\begin{figure}[!t]
\centering
\includegraphics[height=2.3in,clip,keepaspectratio]{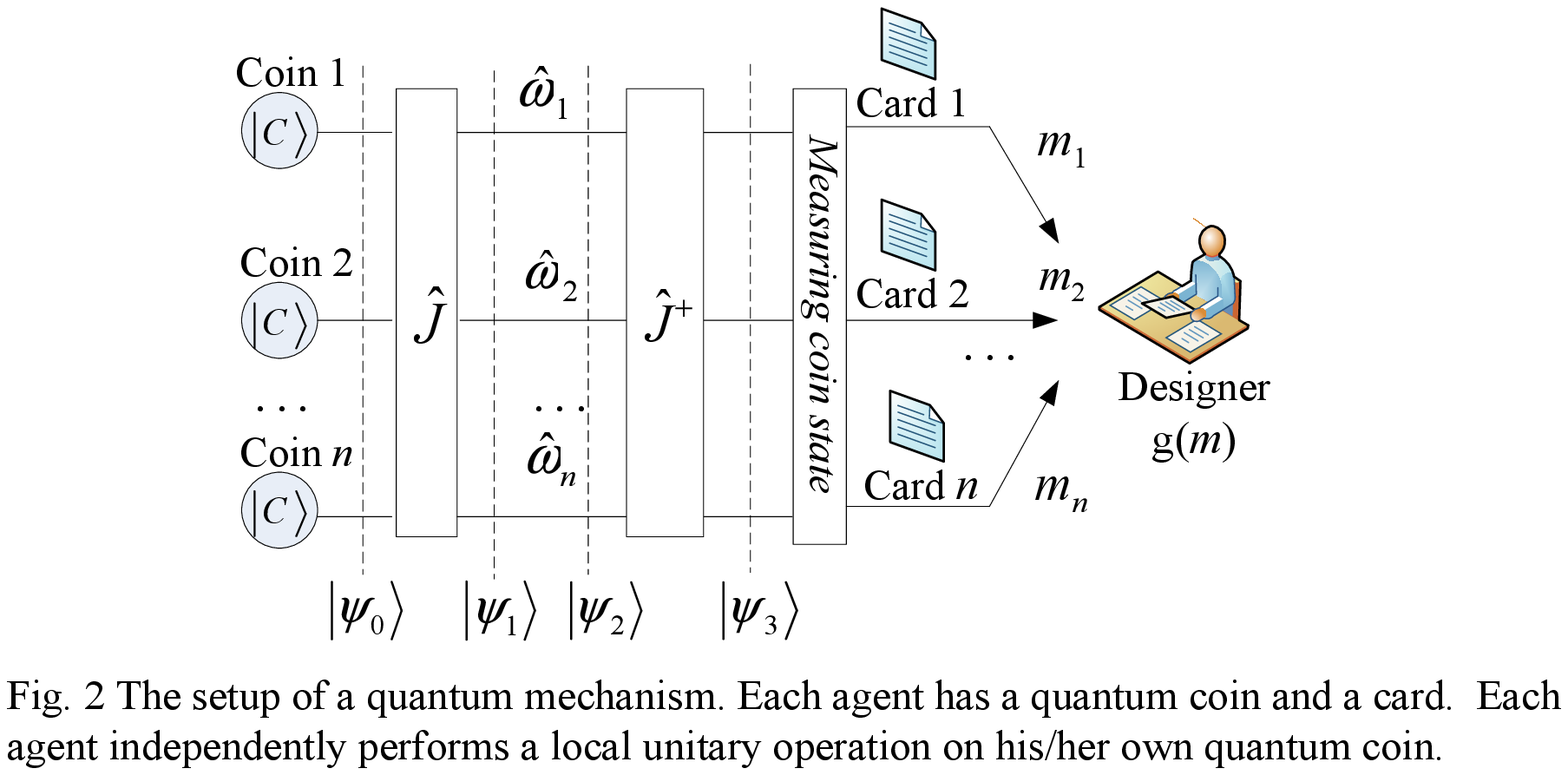}
\end{figure}

Step 1: Nature selects a state $t\in \mathcal {T}$ and assigns $t$
to the agents. The state of every quantum coin is set as
$|C\rangle$. The initial state of the $n$ quantum coins is
$|\psi_{0}\rangle=\underbrace{|C\cdots CC\rangle}\limits_{n}$.\\
Step 2: In state $t$, if all agents agree that the social choice
rule $F$ is ``bad'', i.e., there exist $\hat{t}\in \mathcal {T}$,
$\hat{t}\neq t$, $\hat{a}\in F(\hat{t})$ such that
$\hat{a}\succeq^{t}_{i}a\in F(t)$ for every $i\in N$, and
$\hat{a}\succ^{t}_{j}a\in F(t)$ for at least one $j\in N$, then go
to Step 4.\\
Step 3: Each agent $i$ sets
$c_{i}=((a_{i},t_{i},z_{i}),(a_{i},t_{i},z_{i}))$ (where $a_{i}\in
A$, $t_{i}\in\mathcal {T}$, $z_{i}\in \mathbb{Z}_{+}$),
$\hat{\omega}_{i}=\hat{I}$. Go to Step 7.\\
Step 4: Each agent $i$ sets
$c_{i}=((\hat{a},\hat{t},0),(a_{i},t_{i},z_{i}))$. Let $n$ quantum
coins be entangled by $\hat{J}$. $|\psi_{1}\rangle=\hat{J}|C\cdots
CC\rangle$.\\
Step 5: Each agent $i$ independently performs a local unitary
operation $\hat{\omega}_{i}$ on his/her own quantum coin.
$|\psi_{2}\rangle=[\hat{\omega}_{1}\otimes\cdots\otimes\hat{\omega}_{n}]\hat{J}|C\cdots
CC\rangle$.\\
Step 6: Let $n$ quantum coins be disentangled by $\hat{J}^{+}$.
$|\psi_{3}\rangle=\hat{J}^{+}[\hat{\omega}_{1}\otimes\cdots\otimes\hat{\omega}_{n}]\hat{J}|C\cdots
CC\rangle$.\\
Step 7: The device measures the state of $n$ quantum coins and sends
$card(i,0)$ (or $card(i,1)$) as $m_{i}$ to the designer if the state
of quantum coin $i$ is $|C\rangle$ (or $|D\rangle$).\\
Step 8: The designer receives the overall message
$m=(m_{1},\cdots,m_{n})$ and let the final outcome
$\hat{G}(\hat{s})=g(m)$ using rules (1), (2) and (3) defined in the
Maskin's mechanism. END.

Note that if $\hat{\Omega}_{i}$ is restricted to be $\{\hat{I},
\hat{D}_{n}\}$, then $\hat{\Omega}_{i}$ is equivalent to \{\emph{Not
flip}, \emph{Flip}\}. In this way, a quantum mechanism is
degenerated to a physical mechanism.

Given $n$ ($n\geq 3$) agents, consider the pay-off to the $n$th
agent, we denote by $\$_{C\cdots CC}$ the expected pay-off when all
agents choose $\hat{I}$ (the corresponding collapsed state is
$|C\cdots CC\rangle$), and denote by $\$_{C\cdots CD}$ the expected
pay-off when the $n$th agent chooses $\hat{D}_{n}$ and the first
$n-1$ agents choose $\hat{I}$ (the corresponding collapsed state is
$|C\cdots CD\rangle$). $\$_{D\cdots DD}$ and $\$_{D\cdots DC}$ are
defined similarly. Unlike Flitney and Hollenberg's requirements on
the pay-offs, for the case of quantum mechanism, the requirements on
the pay-offs are described as condition $\lambda$:\\
(i) $\lambda_{1}$: Given  a state $t$ and an SCR $F$, there exist
$\hat{t}\in \mathcal {T}$, $\hat{t}\neq t$, $\hat{a}\in F(\hat{t})$
such that $\hat{a}\succeq^{t}_{i}a\in F(t)$ for every $i\in N$,
$\hat{a}\succ^{t}_{j}a\in F(t)$ for at least one $j\in N$, and the
number of agents that encounter a preference change around $\hat{a}$
in going from state $\hat{t}$ to $t$ is larger than unity. Denote by
$l$ the number of these agents. Without loss of generality, let
these $l$ agents be the last $l$ agents among $n$ agents.\\
(ii) $\lambda_{2}$: Consider the pay-off to the $n$th agent,
$\$_{C\cdots CC}>\$_{D\cdots DD}$, i.e., he/she prefers the expected
payoff of a certain outcome (generated by rule 1) to the expected
payoff of an uncertain outcome (generated by rule 3).\\
(iii) $\lambda_{3}$: Consider the pay-off to the $n$th agent,
$\$_{C\cdots CC}>\$_{C\cdots
CD}[1-\sin^{2}\gamma\sin^{2}(\pi/l)]+\$_{D\cdots
DC}\sin^{2}\gamma\sin^{2}(\pi/l)$.

\textbf{Proposition 2:} For $n\geq3$, given a state
$t\in\mathcal{T}$ and a ``bad'' SCR $F$ (from the viewpoint of
agents) that is monotonic and satisfies no-veto, by virtue of a
quantum mechanism $\Gamma^{Q}=((\hat{S}_{i})_{i\in N},\hat{G})$,
agents satisfying condition $\lambda$ can combat the ``bad'' SCR
$F$, i.e., there exists $\hat{s}\in\mathcal{N}(\Gamma^{Q},t)$ such
that $\hat{G}(\hat{s})\notin F(t)$.

\textbf{Proof:} Given a state $t$ and a ``bad'' SCR $F$, since
condition $\lambda_{1}$ is satisfied, then there exist $\hat{t}\in
\mathcal {T}$, $\hat{t}\neq t$, $\hat{a}\in F(\hat{t})$ such that
$\hat{a}\succeq^{t}_{i}a\in F(t)$ for every $i\in N$,
$\hat{a}\succ^{t}_{j}a\in F(t)$ for at least one $j\in N$, and the
number of agents that encounter a preference change around $\hat{a}$
in going from state $\hat{t}$ to $t$ is larger than unity, i.e.,
$l\geq 2$. Let these $l$ agents be the last $l$ agents among $n$
agents. Hence, the quantum mechanism enters Step 4. Each agent $i$
sets $c_{i}=((\hat{a},\hat{t},0),(a_{i},t_{i},z_{i}))$. Let
$c=(c_{1},\cdots,c_{n})$.

 Consider the pay-off to the $n$th agent
(denoted as \emph{Laura}), when she plays
$\hat{\omega}(\theta,\phi)$ while the first $n-l$ agents play
$\hat{I}$ and the middle $l-1$ agents play
$\hat{C}_{l}=\hat{\omega}(0,\pi/l)$, according to Ref. 9,
\begin{align*}
  \langle\$_{Laura}\rangle=&\$_{C\cdots CC}\cos^{2}(\theta/2)[1-\sin^{2}\gamma
  \sin^{2}(\phi-\pi/l)]\\
  +&\$_{C\cdots CD}\sin^{2}(\theta/2)[1-\sin^{2}\gamma
  \sin^{2}(\pi/l)]\\
  +&\$_{D\cdots DC}\sin^{2}(\theta/2)\sin^{2}\gamma \sin^{2}(\pi/l)\\
  +&\$_{D\cdots DD}\cos^{2}(\theta/2)\sin^{2}\gamma
  \sin^{2}(\phi-\pi/l)
\end{align*}
Since condition $\lambda_{2}$ is satisfied, then $\$_{C\cdots
CC}>\$_{D\cdots DD}$, $Laura$ chooses $\phi=\pi/l$ to minimize
$\sin^{2}(\phi-\pi/l)$. As a result,
\begin{align*}
  \langle\$_{Laura}\rangle=&\$_{C\cdots CC}\cos^{2}(\theta/2)\\
  +&\$_{C\cdots CD}\sin^{2}(\theta/2)[1-\sin^{2}\gamma
  \sin^{2}(\pi/l)]\\
  +&\$_{D\cdots DC}\sin^{2}(\theta/2)\sin^{2}\gamma \sin^{2}(\pi/l)
\end{align*}
Since condition $\lambda_{3}$ is satisfied, then $Laura$ prefers
$\theta=0$, which leads to $\langle\$_{Laura}\rangle=\$_{C\cdots
CC}$. In this case,
$\hat{\omega}_{Laura}(\theta,\phi)=\hat{\omega}(0,\pi/l)=\hat{C}_{l}$.

By symmetry, in Steps 4 and 5, if the $n$ agents choose
$\hat{s}^{*}=(\hat{\omega}^{*}, c)$, where
$\hat{\omega}^{*}=(\hat{I},\cdots,\hat{I},\hat{C}_{l},
\cdots,\hat{C}_{l})$ (the first $n-l$ agents choose $\hat{I}$, the
other $l$ agents choose $\hat{C}_{l}$), then
$\hat{s}^{*}\in\mathcal{N}(\Gamma^{Q},t)$. In Step 7, the
corresponding collapsed state of $n$ quantum coins is $|C\cdots
CC\rangle$ and $m_{i}=(\hat{a},\hat{t},0)$ for each agent $i\in N$.
Consequently, in Step 8, $\hat{G}(\hat{s}^{*})=g(m)=\hat{a}\notin
F(t)$. $\quad\quad\quad\quad\quad\quad\quad\quad\quad\quad\quad\quad
\quad\quad\quad\quad\quad\quad\quad\quad\quad\quad\quad\square$

Let us reconsider Example 1. The quantum mechanism enters Step 4
when the true state is $t_{2}$. Since both $Apple$ and $Lily$
encounter a preference change around $a_{1}$ in going from state
$t_{1}$ to $t_{2}$, condition $\lambda_{1}$ is satisfied.
$c_{Apple}=((a_{1},t_{1},0),(a_{4},*,*))$,
$c_{Lily}=((a_{1},t_{1},0),(a_{3},*,*))$,
$c_{Cindy}=((a_{1},t_{1},0),(a_{1},t_{1},0))$. Let $Cindy$ be the
first agent. For any agent $i\in\{Apple, Lily\}$, let her be the
last agent. Consider the pay-off to the third agent, suppose
$\$_{CCC}=3$ (the corresponding outcome is $a_{1}$), $\$_{CCD}=5$
(the corresponding outcome is $a_{4}$ if $i=Apple$, and $a_{3}$ if
$i=Lily$), $\$_{DDC}=0$ (the corresponding outcome is $a_{3}$ if
$i=Apple$, and $a_{4}$ if $i=Lily$), $\$_{DDD}=1$ (the corresponding
outcome is $a_{3}/a_{4}$). Hence, condition $\lambda_{2}$ is
satisfied, and condition $\lambda_{3}$ becomes:
$3\geq5[1-\sin^{2}\gamma\sin^{2}(\pi/2)]$. If
$\sin^{2}\gamma\geq0.4$, condition $\lambda_{3}$ is satisfied.
According to Proposition 2, the message corresponding to
$\hat{s}^{*}\in\mathcal{N}(\Gamma^{Q},t)$ is
$m=(m_{1},m_{2},m_{3})$, where $m_{1}=m_{2}=m_{3}=(a_{1},t_{1},0)$.
Consequently, $\hat{G}(\hat{s}^{*})=g(m)=a_{1}\notin
F(t)=\{a_{2}\}$.

To help the reader understand the aforementioned result, let the SCR
in Table 1 be ``No smoking''. Let $a_{1}$ and $a_{2}$ denote
``Smoke'' and ``Drink'' respectively, then everybody prefers smoking
to drinking in state $t_{2}$. According to the traditional theory of
mechanism design, the ``No smoking'' SCR can always be Nash
implemented because it is monotonic and satisfies no-veto. However,
by virtue of quantum strategies, the agents can combat the ``No
smoking'' SCR!

\textbf{Remark:} In Maskin and Sj\"{o}str\"{o}m  $^{4}$, the authors
used a modulo game instead of the integer game. The rule 3 is
replaced by ``(3) In all other cases, $g(m)=a_{j}$, for $j\in N$
such that $j=(\sum_{i\in N}z_{i})(\mbox{mod }n)$''. Similar to
aforementioned analysis, it can be derived that the results of this
paper still hold.

\section{Conclusion}
In conclusion, this paper considers what will happen if agents can
use quantum strategies in the theory of mechanism design. Two
results are obtained: (i) We find that the success of the Maskin's
mechanism is built on an underlying Prisoner's Dilemma. (ii) Under
the classical circumstance, if an SCR is monotonic and satisfies
no-veto, then no matter whether it is ``bad'' or not (from the
viewpoint of the agents), it can be Nash implemented. However, we
find that when the additional condition $\lambda$ is satisfied, an
original Nash implementable ``bad'' SCR will no longer be Nash
implementable in the context of a quantum domain.

van Enk and Pike $^{11}$ pointed out that in quantum games, quantum
strategies simply constructed a new game and solved it, not the
original game. However, from the viewpoint of the designer, the
interface between agents and the designer in the quantum mechanism
is the same as that in the Maskin's mechanism. Therefore, from the
viewpoint of agents, quantum mechanism helps them combat ``bad''
social choice rules specified by the designer.

\section*{Acknowledgments}

The author is very grateful to Ms. Fang Chen, Hanyue Wu
(\emph{Apple}), Hanxing Wu (\emph{Lily}) and Hanchen Wu
(\emph{Cindy}) for their great support.

\vspace*{-6pt}   

\section*{References}
\vspace*{-5pt}   

\myitem J. von Neumann and O. Morgenstern, \emph{Theory of Games and
Economic Behavior} (Princeton University Press, USA, 1944).

\myitem R. Serrano, \emph{SIAM Review} \textbf{46} (2004) 377.

\myitem E. Maskin, \emph{Rev. Econom. Stud.} \textbf{66} (1999) 23.

\myitem E. Maskin and T. Sj\"{o}str\"{o}m, Implementation theory, in
\emph{Handbook of Social Choice and Welfare}, Vol. 1, eds. K. J.
Arrow, A. Sen, K. Suzumura (Elsevier Science, New York, 2002).

\myitem J. Eisert, M. Wilkens and M. Lewenstein, \emph{Phys. Rev.
Lett.} \textbf{83} (1999) 3077.

\myitem D. Meyer, \emph{Phys. Rev. Lett.} \textbf{82} (1999) 1052.

\myitem S.C. Benjamin and P.M. Hayden, \emph{Phys. Rev. A}
\textbf{64} (2001) 030301(R) .

\myitem J. Du, H. Li and X. Xu \emph{et al}, \emph{Phys. Lett. A}
\textbf{302} (2002) 229.

\myitem A.P. Flitney and L.C.L. Hollenberg, \emph{Phys. Lett. A}
\textbf{363} (2007) 381.

\myitem H. Guo, J. Zhang and G.J. Koehler, \emph{Decision Support
Systems} \textbf{46} (2008) 318.

\myitem S. J. van Enk and R. Pike, \emph{Phys. Rev. A} \textbf{66}
(2002) 024306.

\end{document}